# Penetration depth of low-coherence enhanced backscattered light in sub-diffusion regime

**Hariharan Subramanian, Prabhakar Pradhan, Young L. Kim and Vadim Backman**

*Biomedical Engineering Department, Northwestern University, Evanston, IL 60208.*

**Abstract:**

The mechanisms of photon propagation in random media in the diffusive multiple scattering regime have been previously studied using diffusion approximation. However, similar understanding in the low-order (sub-diffusion) scattering regime is not complete due to difficulties in tracking photons that undergo very few scatterings events. Recent developments in low-coherence enhanced backscattering (LEBS) overcome these difficulties and enable probing photons that travel very short distances and undergo only a few scattering events. In LEBS, enhanced backscattering is observed under illumination with spatial coherence length $L_{sc}$ less than the scattering mean free path $l_s$. In order to understand the mechanisms of photon propagation in LEBS in the sub-diffusion regime, it is imperative to develop analytical and numerical models that describe the statistical properties of photon trajectories. Here we derive the probability distribution of penetration depth of LEBS photons and report Monte Carlo numerical simulations to support our analytical results. Our results demonstrate that, surprisingly, the transport of photons that undergo low-order scattering events has only weak dependence on the optical properties of the medium ($l_s$ and anisotropy factor $g$) and strong dependence on the spatial coherence length of illumination, $L_{sc}$ relative to those in the diffusion regime. More importantly, these low order scattering photons typically penetrate less than $l_s$ into the medium due to low spatial coherence length of illumination and their penetration depth is proportional to the one-third power of the coherence volume (i.e. $[l_s \pi L_s^2]^{1/3}$).

# I. INTRODUCTION

Most biological tissues are multilayered systems that require depth-selective measurements to obtain clinically useful information [1-6]. Currently, a number of optical techniques based on backscattered light are under development for such depth-selective tissue characterization or imaging. In order to exploit an optical technique in a biomedical setting, a proper knowledge of the photon trajectories within the sample before being backscattered is essential. This information can be characterized by the distribution of photons at different depths, herein called penetration depth distribution, which provides information about the probability that a photon penetrate a certain depth before being detected. The penetration depth distribution, in turn, can be conveniently characterized by the effective penetration depth (depth corresponding to the peak of the probability distribution curve). Several groups have used numerical and analytical models to study the penetration depth of backscattering photons in tissues in a multiple scattering medium [7-10]. In particular, Weiss *et al.* used lattice random walk models to obtain the statistical properties of the penetration depth of photons emitted from a bulk tissue [7, 8]. The depth distribution of photons in a random scattering medium with the thickness of approximately 10 transport mean free paths ($l_s^*$) was calculated by Durian [9]. Recently, Zaccanti *et al.* [10] derived analytical expressions for the time resolved probability of photons penetrating a certain depth in a diffusive medium, before being re-emitted. Although the penetration depth of photons has been well studied in a diffusive multiple scattering regime, similar understanding in the low-order scattering regime is not complete. This is in part due to the difficulties in collecting photons that undergo only a few scattering events. Recently, we have developed low-coherence enhanced backscattering (LEBS), which spatially filters longer traveling photons and collects only photons that travel very short distances and undergo a few orders of scattering. In this paper, we use Monte Carlo numerical simulation to study the propagation of photons that contribute to LEBS; and also report development of a corresponding analytical model to describe the penetration depth distribution and effective penetration depth of these photons.

Enhanced backscattering (EBS, also known as coherent backscattering) is a phenomenon in which coherent photons traveling along exact time-reversed paths interfere constructively to produce an enhanced intensity peak in the directions close to backscattering. Therefore, theoretically, the intensity of the EBS peak in the backward direction can be as high as twice the diffused background. Typically, the angular width of the EBS peak is proportional to $\lambda/l_s^*$, where $\lambda$ is the wavelength of light and $l_s^*$ is the transport mean free path length [11, 12]. Although the EBS phenomenon has been extensively studied in a variety of non-biological media [13-20], the investigation of EBS in biological tissue has been extremely limited [21-23]. A biological tissue is a weakly scattering medium ($l_s^* \gg \lambda$) with $l_s^*$ ranging between 0.5 to 2mm. The investigation of EBS in such weakly scattering media has been exceedingly difficult due to very small widths of EBS peaks (e.g., $w_{hm} \sim 0.001°$ for $l_s^* \sim 1$ mm). On the contrary, low coherence EBS (LEBS) overcomes all of the major limitations that have prevented the widespread application of EBS in tissue optics [24, 25]. The LEBS peak is obtained by combining the EBS measurements with low spatial coherence, broadband illumination. In our previous studies we showed that low spatial coherence illumination (spatial coherence length $L_{sc} < l_s \ll l_s^*$) behaves as a spatial filter that dephases the conjugated time-reversed paths outside the spatial coherence area and thus rejects longer path lengths [24, 25]. This gives rise to EBS peaks that are broadened by more than two orders of magnitude compared to the width of conventional EBS peaks, facilitating their experimental measurements [24, 25]. We have also previously shown that LEBS opens up the feasibility of studying tissue optical properties at a selected depth: LEBS can selectively probe short traveling photons from the top tissue layer (50-100 µm, e.g. mucosa and epithelium,) by rejecting long traveling photons from the underlying (stromal/connective) tissue. Finally, we have shown that LEBS spectroscopy can reliably identify the earliest precancerous alterations in the colon and pancreas [24, 25] using LEBS properties such as its spectral and angular distributions.

As discussed above, in LEBS, low-spatial coherence illumination acts as a spatial filter that rejects longer traveling photons. Therefore, the penetration depth of LEBS photons can be controlled externally by changing the spatial coherence length of illumination, $L_{sc}$. In order to increase the sensitivity of LEBS measurements to specific tissue depths it is important to know the relationship between $L_{sc}$ and the depth of penetration of LEBS photons. The penetration depth in turn is characterized by studying the penetration depth distribution of the short traveling LEBS photons $p(z)$ (where, $p(z)$ is the probability of photons to penetrate a certain depth $z$ before being detected), and their effective penetration depths ($z_{mp}$). Unlike the long traveling photons, the short traveling LEBS photons ($r < l_s \ll l_s^*$, where $r$ is the radial distance at which photons emerge) typically undergo very few scattering events; hence, the numerical model and analytical expressions addressed within the diffusion approximation for $r \gg l_s^* > l_s$ cannot be used to study the mechanisms of photons propagation in LEBS. Therefore, we used numerical

Monte Carlo (MC) simulations to model low-order scattered photons in order to study the mechanisms of photon propagation as a function of coherence length $L_{sc}$, anisotropy coefficient $g$ and scattering mean free path length $l_s$. We also developed an analytical model for $p(z)$ and $z_{mp}$ of LEBS photons, i.e. photons exiting at radial distances $r < l_s \ll l_s^*$ with $L_{sc} < l_s \ll l_s^*$.

In order to study the penetration depth of LEBS photons, we perform the following: First, we use the numerical Monte Carlo simulation to calculate the penetration depth distribution $p(z)$ and the effective penetration depth ($z_{mp}$) of the photons that form the LEBS peak. We study this for different coherence lengths ($L_{sc}$) of illumination, and for media with different scattering mean free paths ($l_s$) and anisotropies ($g$). Second, we use a double scattering analytical model to develop analytical expressions for $p(z)$ and $z_{mp}$, and show that the analytical expressions of $p(z)$ and $z_{mp}$ compare well with the corresponding numerical results for the parameter regime when $L_{sc} < l_s \ll l_s^*$. Finally, we demonstrate that both $p(z)$ and $z_{mp}$ of the exiting photons in this regime ($L_{sc} < l_s \ll l_s^*$) exhibit a priori surprising behavior, that is, only weak dependence on optical properties ($l_s$ and $g$) and strong dependence on $L_{sc}$ relative to diffusive regime. This result is in contrary to the general understanding of the properties of $p(z)$ and $z_{mp}$ observed in the diffusive regime.

## II. METHODS

A. Numerical model using Monte Carlo simulation

Monte Carlo (MC) simulations have been commonly used to model photon transport in biological media [26, 27], and also to model EBS phenomenon indirectly [19, 28-30]. As it is challenging to simulate the time-reversal of photons and its interference effects explicitly using MC methods, EBS angular profiles $I_{EBS}(\theta)$ are generally calculated by using the fundamental relationship between $I_{EBS}(\theta)$ and the radial distribution *p(r)* with *p(r)* being the probability of photons to emerge from the surface at a radial distance *r*. That is, $I_{EBS}(\theta)$ is an integral transform of the radial intensity distribution of *p(r)*, where, in turn, *p(r)* is obtained from the MC simulations [12]:

$$I_{EBS}(\vec{q}_\perp) \propto \int r p(r) \exp(i 2\pi r \sin\theta / \lambda) dr \tag{1}$$

where *r* is the radial distance at which the photons emerge and $\vec{q}_\perp$ is the projection of the wave vector onto the plane orthogonal to the backward direction.

In order to explore the depth of penetration of LEBS photons and its dependence on the optical properties of a medium, we use a MC simulation method developed by Wang *et al*. [26]. Although, the propagation of photons and its dependence on optical properties have been well studied in the diffusion regime using MC simulations, here we use MC to study the low order scattering, particularly when the photons undergo minimum of double scattering events and then exit within a narrow radial distance ($r < l_s \ll l_s^*$). The double scattering is of significance because in EBS the minimum number of scattering event is double scattering. The single scattering events contribute only to the incoherent baseline and not to the EBS peak formation. We have recently demonstrated a direct experimental evidence that double scattering is the minimal scattering event necessary to generate an EBS peak in a discrete random medium [31]. In this study, we also showed that LEBS isolates double scattering from higher order scattering when $L_{sc}$ is on the order of the scattering mean free path $l_s$ of light in the medium ($l_s = l_s^*(1-g)$).

Description of the MC simulation is given in detail elsewhere [26, 27]. In brief, we launch an infinitely narrow photon beam consisting of $10^{10}$ photon packets into a homogeneous disordered single layered medium with thickness much greater than the spatial extent of the photon distribution (thickness of the medium = 50mm). We vary the scattering mean free path $l_s$ between 50-500 μm and the anisotropy factor $g$ between 0.7-0.9. We assume absorption to be negligible (absorption length, $l_a$ = 1000 cm). We record the trajectories of all photons that undergo two and higher order scattering events, and exit the sample at an angle < 3° from the direction of backscattering. We obtain the penetration depth distribution of photons in the axial 'z' direction ($p(z)$) using a two-dimensional grid system whose grid separations in the $r$ and $z$ directions were $\delta_r$ = 2 μm and $\delta_z$ = 5 μm, respectively, with the total number of grids $N_r = N_z$ = 1000. Furthermore, to account for the number of scattering events ($n_s$), we setup a separate two-dimensional grid system with $\delta_r$ = 2 μm, interval between scattering events $\delta_{n_s} = 1$, total number of grids $N_r$ = 1000 and the total number of scattering events $N_{n_s} = 500$, respectively. Therefore, we can obtain the scattering distribution $p(n_s)$, penetration depth distribution $p(z)$ and the effective penetration depth $z_{mp}$ as a function of radial distance $r$.

We also calculate $p(n_s)$, $p(z)$, and $z_{mp}$ as a function of spatial coherence length $L_{sc}$, by incorporating the effect of low spatial coherence illumination on EBS in the numerical model. In this case, the angular profile of LEBS $I_{LEBS}(\theta)$ can be expressed as [25],

$$I_{LEBS}(\theta) = \int_0^\infty C(r)rp(r)\exp(i2\pi r\theta/\lambda)\,dr, \qquad (2)$$

where $C(r) = |2J_1(r/L_{sc})/(r/L_{sc})|$ is the degree of spatial coherence of illumination with the first order Bessel function $J_1$ [32]. As $C(r)$ is a decay function of $r$, it acts as a spatial filter allowing only photons emerging within its effective coherence area ($\sim L_{sc}^2$) to contribute to $p(r)$. Therefore, we can obtain $p(z)$ and $z_{mp}$ as a function of $r$ or $L_{sc}$.

The following section discusses in detail the derivation of the analytical expressions of $p(z)$ and $z_{mp}$ from a double scattering analytical model, and its comparison with the results of our numerical simulations.

B. Analytical derivation of $p(z)$, and $z_{mp}$

We derive the expressions for $p(z)$ and $z_{mp}$ of photons that contribute to LEBS peak on the basis of a double-scattering analytical model of backscattering photons. Previous experimental studies [31] and the numerical results of the Monte Carlo simulations, which will be discussed in detail below (Section III.A), demonstrate that LEBS peaks from a low spatial coherence illumination are mainly generated by the photons that predominantly undergo double scattering events. Hence, we use the double scattering analytical model to derive the expressions for $p(z)$ and $z_{mp}$ and to verify our results from the numerical simulations.

*1. $p(z)$ and $z_{mp}$ as a function of radial distance r:*

The probability of radial distribution of photons exiting a medium $p(r)$ due to double scattering events can be expressed as [33],

$$p(r) = \int_0^\infty \int_0^\infty \frac{dz' dz''}{r^2 + (z''-z')^2} \exp[-\mu_s(\sqrt{r^2 + (z''-z')^2} + z' + z'')]\mu_s F(\theta)\mu_s F(\pi - \theta), \quad (3)$$

where $r$ is the radial distance at which photons emerge, $z'$ and $z''$ are the vertical distances from the surface to the scatterers, $F(\theta)$ is the phase function of single scattering with $\theta = \tan^{-1}(r/(z''-z'))$, and $\mu_s (\equiv 1/l_s)$ is the scattering coefficient. A schematic picture of the scattering geometry is shown in Fig. 1. In our study, we use the Henyey-Greenstein scattering phase function,

$$F(\theta) = \frac{1}{4\pi} \frac{1-g^2}{\left(1+g^2 - 2g\cos\theta\right)^{3/2}}. \tag{4}$$

To obtain the expressions of $p(z)$ and $z_{mp}$ of a double scattering photon from Eq. (2), we perform the following: We define a new variable $z = z''-z'$ and $z''' = z''+z'$. The coordinate system, in the above double scattering model, can be transformed to a $zz'''$ coordinate system, using a Jacobian transformation. We then approximate the penetration depth of the double scattering events as $\sim z$, as one of the scattering events occurs much closer to the surface of the medium than the other scattering event when the exit distances $r$ of the majority of photons are restricted due to the finite value of $L_{sc}$ ($r, L_{sc} < l_s \ll l_s^*$). Indeed, Fig. 1 illustrates that in order for the photons to undergo double scattering events within a small $r$, one of the scattering events must occur very close to the surface of the medium ($z' \approx 0$) (approximation validated in Section III.B). Therefore, the double scattering expression can be rewritten as,

$$p(r) = \int_0^\infty \int_0^\infty \frac{2 dz dz'''}{r^2 + z^2} \exp[-\mu_s(\sqrt{r^2+z^2} + z''')] \mu_s F(\theta) \mu_s F(\pi - \theta). \tag{5}$$

Integrating over $z'''$ in Eq. (5) we obtain,

$$p(r) = \int_0^\infty p(r,z) dz = \int_0^\infty \frac{2}{\mu_s} \frac{dz}{r^2 + z^2} \exp[-\mu_s(\sqrt{r^2+z^2})] \mu_s F(\theta) \mu_s F(\pi - \theta). \tag{6}$$

From Eq. (6) it follows that for a given $r$, the penetration depth distribution $p(z)$ can be written as,

$$p(z|r) = \frac{2}{\mu_s} \frac{1}{r^2+z^2} \exp[-\mu_s(\sqrt{r^2+z^2})]\mu_s F(\theta)\mu_s F(\pi-\theta) . \quad (7)$$

Substituting $\theta = \tan^{-1}(r/z)$, the phase function Eq. (4) can be rewritten as,

$$F(\theta) = \frac{1}{4\pi} \frac{1-g^2}{\left(1+g^2-2g\frac{z}{\sqrt{r^2+z^2}}\right)^{3/2}} . \quad (8)$$

Because the phase function F is mostly uniform around the backward direction, i.e. $\theta \sim \pi$, we approximate, $F(\pi-\theta,\lambda) \approx 1$. Then the penetration depth distribution at a given $r$, $p(z|r)$ becomes,

$$p(z|r) = \frac{\mu_s}{2\pi} \frac{1}{r^2+z^2} \exp[-\mu_s(\sqrt{r^2+z^2})] \frac{1-g^2}{\left(1+g^2-2g\frac{z}{\sqrt{r^2+z^2}}\right)^{3/2}} . \quad (9)$$

Eq. (9) is the depth distribution of photons that undergo double scattering events and exit the medium in the backward direction at radial distances $r < l_s \ll l_s^*$.

The effective penetration depth $z = z_{mp}$ is the solution of the following equation.

$$\left. \frac{dp(z)}{dz} \right|_{z=z_{mp}} = 0 . \quad (10)$$

From the Eqs. (9) and (10) we obtain,

$$z_{mp}^3 + \left[\frac{2}{\mu_s}\right] z_{mp}^2 + \left[r^2 + \frac{r^2 g}{1+g^2-2g}\right] z_{mp} + \left[\frac{r^2}{\mu_s} - \frac{r^2 g}{\mu_s(1+g^2-2g)}\right] = 0 \quad (11)$$

Solving the above cubic equation (Eq. (11)) for $z_{mp}$, we obtain the exact solution for the effective penetration depth $z_{mp}$ of double scattering photons:

$$z_{mp}(r|g,\mu_s) = \frac{2}{3\mu_s} \frac{1}{(1-g)^2} \left[ \frac{\left[ B(r,g,\mu_s) + \sqrt{B^2(r,g,\mu_s) + 4A^3(r,g,\mu_s)} \right]^{1/3}}{2^{4/3}} + \right. \\ \left. - \frac{A(r,g,\mu_s)}{2^{2/3} \times \left[ B(r,g,\mu_s) + \sqrt{B^2(r,g,\mu_s) + 4A^3(r,g,\mu_s)} \right]^{1/3}} \right] - \frac{2}{3\mu_s}, \quad (12)$$

where

$$A(r,g,\mu_s) = (1-g)^4 \left[ -4 + \frac{3gr^2\mu_s^2}{(1-g)^2} \right], \text{ and}$$

$$B(r,g,\mu_s) = (1-g)^6 \left[ -16 + \frac{45gr^2\mu_s^2}{(1-g)^2} \right].$$

The dependence of $z_{mp}$ on $r$ (from Eq. (12) ) for $\mu_s r \approx 0$ is approximately linear. Here we are interested in $z_{mp}$ in the regime relevant for LEBS: $r/l_s < 1$, $(1-g)^2 (\mu r)^2 \to 0$, and $(g)^2 (\mu r)^2 \sim 1$. To see the leading behavior of $z_{mp}$ in this regime, we expand the right side of Eq. (12) in terms of $A$ and $B$, when $A/B \ll 1$ and obtain $z_{mp} \propto \left( 2/(3\mu_s(1-g)^2) \right) \left[ 2B(r,g,\mu_s) \right]^{1/3}$. This can be re-written as:

$$z_{mp}(r|g,l_s) \propto \frac{g^{1/3}}{(1-g)^{2/3}} \left[ l_s \pi r^2 \right]^{1/3}. \quad (13)$$

The above equation implies that the effective penetration depth of a LEBS photon is proportional to the (1/3) power of the volume of a virtual cylinder whose area is formed by a circle of radius $r$ and height $l_s$. In the case for biological tissue (*i.e.*, $g \sim 1$), Eq. (13) can be rewritten as, $\frac{z_{mp}}{l_s} \propto \left[ \frac{r}{l_s^{**}} \right]^{2/3}$, where $l_s^{**} = l_s(1-g)$. This provides a critical value of $r$ in the units of $l_s$ for the double scattering regime; *i.e.,* for $r < l_s^{**}$, double scattering events dominate compared to higher order scattering events.

*2. p(z) and $z_{mp}$ as a function of spatial coherence length $L_{sc}$:*

To calculate the dependence of $p(z)$ and $z_{mp}$ on coherence length $L_{sc}$, we first weight the Eqs. (9) and (12) by the coherence function $C(r, L_{sc})$, and integrate over $r$:

$$p(z | L_{sc}) = \int_{r=0}^{\infty} p(z | r) C_{L_{sc}}(r) dr, \tag{14}$$

$$z_{mp}(L_{sc} | g, \mu_s) = \int_{r=0}^{\infty} z_{mp}(r | g, \mu_s) C_{L_{sc}}(r) dr. \tag{15}$$

Eqs. (14) and (15) represent the analytical expressions for the penetration depth distribution, and effective penetration depth of photons that predominantly undergo double scattering events in LEBS.

Under low-coherence illumination with spatial coherence length $L_{sc} < l_s << l_s^*$: $\mu_s r < 1$ and $C(r, L_{sc}) dr \sim dr/L_{sc}$. In this low-coherence regime, integration in Eq. (15) can be performed analytically.

$$z_{mp}(L_{sc} | g, l_s) \sim \int_0^{l_{sc}} z_{mp}(r | g, L_s) C(r, L_{sc}) dr \quad \propto \quad \frac{g^{1/3}}{(1-g)^{2/3}} \left[ l_s \pi L_{sc}^2 \right]^{1/3} \tag{16}$$

Eq. (16) implies that the effective penetration depth of a LEBS photon is proportional to the 1/3 power of an effective coherence volume in a large parameter space. For example, at g = 0.7 and $l_s$ = 100 μm, the plot of $\log(z_{mp})$ versus $\log(\pi l_s L_{sc}^2)$ has a slope of 0.28 (~1/3).

## III. RESULTS AND DISCUSSIONS

A. Scattering and penetration depth distribution - numerical studies

The probability with which a photon scatters, *p(n_s)*, and the depth to which it penetrates, *p(z)*, before it exits the medium at radial distances $r < l_s \ll l_s^*$ is discussed in this section. As stated throughout this paper, we consider a low-coherence regime: $L_{sc} < l_s \ll l_s^*$. We performed numerical simulations using MC (Section II.A) for media with different optical properties ($l_s$ = 50-500 µm and g = 0.7-0.9) in order to obtain *p(n_s)* and *p(z)*. As an illustration, here we discuss the results obtained for a medium with $l_s$ = 100 µm and g = 0.9 and 0.7.

Figures 2(a) and 2(b) show *p(n_s)* for photons that exit at $r < l_s \ll l_s^*$ for g = 0.9 and 0.7. For *r* = 5 µm ($r/l_s$ = 0.05), it can be clearly seen that the photons predominantly undergo double scattering events (Fig. 2(a)). This can be seen from a sharp peak in *p(n_s)* for $n_s$ = 2. However, as *r* increases (*r*>25 µm) the probability of collected photons to undergo higher order scattering ($n_s > 2$) increases.

Typically in a medium consisting of small particles (g << 1), photons undergo isotropic scattering and hence penetrate shallower distances than in the medium with large anisotropy factor (g ~ 0.9). As a result, the photons propagating in a sample with small g undergo relatively few scattering events before exiting the sample. This effect can be seen in Fig. 2(b), where the scattering distribution *p(n_s)* is obtained for a medium with g = 0.7. In this case, the shape of *p(n_s)* as a function of $n_s$ is considerably sharper than *p(n_s)* for g = 0.9 (*r* = 5 µm), illustrating that the photons have higher probability of exiting the medium after undergoing double scattering events. It is also interesting to note that for small particles, the probability of two and three scattering events of photons are comparable at *r* = 50 µm.

In the case of LEBS, the coherence area within which a photon exits a medium is controlled by the $L_{sc}$ of the light source. The plots of $p(n_s)$ for three different values of $L_{sc}$ for samples with g = 0.9 and 0.7 are shown in Figs. 2(c) and 2(d). Within a narrow coherence area defined by $L_{sc} < l_s << l_s^*$ (e.g., $L_{sc}$= 5 µm), it can be seen that the majority of the photons experience double scattering while the probability of collecting photons undergoing higher orders of scattering is exponentially low. However, for $L_{sc}$= 50 µm the probabilities of 3 and 4 scattering events are comparable to that of double scattering. These results are critical to the following discussion as they validate our use of the double scattering model to derive the analytical expressions for $p(z)$ and $z_{mp}$ in the low-coherence regime ($L_{sc} < l_s << l_s^*$).

Figure 3 shows numerical simulations of $p(z)$ for two sets of optical properties ($l_s$=100 µm, g=0.9 and $l_s$=100 µm, g=0.7) at different radial distance $r$ ( $r$ = 5 µm, 25 µm, 50 µm ) and spatial coherence length $L_{sc}$ ($L_{sc}$ = 5 µm, 25 µm, 50 µm). For $r$ = 5 µm << $l_s$, the photons typically penetrate a shallow distance into the medium, which is, importantly, less than the scattering mean free path of the medium $l_s$. Also, for a constant g and $l_s$ the penetration depth of the photon increases with increase in the radial distance $r$ at which photons exit the medium (Fig. 3). However, when the results of $p(z)$ are compared to the medium with different optical property (g = 0.7, $l_s$ = 100 µm), the change in $p(z)$ is considerably less significant ( < 5%). This indicates that $p(z)$ is only weakly dependent on the optical properties of the medium for small radial distances $r < l_s << l_s^*$. On the other hand, $p(z)$ shows a strong dependence on $r$, and the shape of $p(z)$ vary significantly (> 50 %) for different $r$. Similarly, the penetration depth distributions at different $L_{sc}$ also show a relatively weak dependence on optical properties, and strong dependence on $L_{sc}$ (Figs. 3(c) and 3(d)). This weak dependence of penetration depth on optical properties for photons exiting at $r, L_{sc} < l_s \ll l_s^*$ was further verified by our numerical simulation for other values of $l_s$ and $g$ of the medium (data not shown). It is interesting to note that within a narrow coherence area defined by $L_{sc} < l_s \ll l_s^*$ (e.g., $L_{sc}$ = 5 µm), the majority of photons penetrate only to a shallow depth ($\sim$ 25 µm $< l_s$). However, as $L_{sc}$ increases, the photons have a higher probability of penetrating deeper into the medium.

From these results, we conclude that the tissue depths that are predominantly sampled by the LEBS photons can be controlled by varying the spatial coherence length of illumination $L_{sc}$, and the resulting penetration depth of the photons is essentially insensitive to the specifics of the tissue optical properties. Also, the LEBS photons typically penetrate a shallow distance which is less than the scattering mean free path $l_s$ of the medium.

B. Comparison of the results of numerical simulations and analytical model

Here we compare the analytical expressions of $p(z)$ and $z_{mp}$ as a function of $r$ (Eq. (9) and Eq. (12)) and $L_{sc}$ (Eq. (14) and Eq. (15)) with the corresponding numerical simulations in the low-coherence regime: $L_{sc} < l_s \ll l_s^*$. As a representative illustration, the analytical and numerical results are shown for a medium with $l_s$ = 100 µm and $g$ = 0.9.

First we validate our hypothesis stated in Section II.B1 that in the double scattering regime when $r, L_{sc} < l_s \ll l_s^*$, the distance from the surface of the medium to one of the scatterers is negligibly small relative to that of the other scatterer, *i.e.* the vertical distance to the deeper scatterer is several orders greater than the other scatterer (either $z' \ll z''$ or $z'' \ll z'$). To validate this hypothesis, we used MC simulations analogous to the one discussed in Section II.A. This time, however, we followed photons that undergo only single scattering. Fig. 4 shows the plots of *p(z)* of photons that undergo single scattering and those undergoing double scattering events for $r = 5$ μm. It is seen that the shape of *p(z)* of a single scattering photon is several times sharper than that of double scattering photons. This confirms that for $r < l_s \ll l_s^*$, the distance to the first scatterer is negligibly smaller than the distance of the second scatterer which is located much deeper within the medium. That is, in order for the photons to undergo double scattering and exit within a narrow radial distance $r < l_s \ll l_s^*$ (e.g., restricted by $L_{sc} < l_s \ll l_s^*$), one of the scattering events must occur close to the surface of the medium. Hence, for a photon undergoing double scattering, the difference in vertical distances between the two scatters can be taken as the penetration depth of this photon. The validation of this assumption will be important for the validation of the *p(z)* obtained using the analytical model with the predictions of numerical simulations.

Fig. 5(a) compares *p(z)* given by Eq. (9) and the one obtained by the numerical model for *r* = 5 μm, 25 μm and 50 μm. As seen in the double scattering regime (i.e., $r, L_{sc} < l_s \ll l_s^*$), the predictions of the analytical model are in good agreement with those of the numerical simulations with root mean square error (RMSE) of less than 0.5 %. Similarly, *p(z)* obtained for $L_{sc}$ = 5 μm, 25 μm and 50 μm indicates that the analytical expression derived from the double scattering model (Eq. (14)) can aptly describe the distributions obtained from the numerical model (RMSE < 0.4 %) (Fig. 5(b)). Even though the numerical model takes into account higher order scattering events, Figs. 5(a) and 5(b) clearly show that for $r, L_{sc} < l_s \ll l_s^*$, the analytical and numerical results are in good agreement. This result further confirms that for $r, L_{sc} < l_s \ll l_s^*$, the photons predominantly undergo double scattering events, and Eq. (9) and Eq. (14) can accurately model the penetration depth distribution of the photons.

Figure 6 compares $z_{mp}$ of LEBS photons predicted by the numerical model and analytical expression (Eq. (15)). Here, $z_{mp}$ is obtained as a function of $L_{sc}$ for two different media with g = 0.7, g = 0.9, and $l_s$ = 100 μm. It can be seen from this plot that in the low-coherence regime ($L_{sc} < l_s \ll l_s^*$), the predictions of $z_{mp}$ by the analytical double-scattering model are in good agreement with those of the numerical simulations for all $L_{sc} < l_s$ (RMSE < 5 μm). This good agreement is due to the fact that, as discussed above, double scattering dominates in this regime. Furthermore, even if a photon undergoes a higher order scattering, the condition $L_{sc} < l_s \ll l_s^*$ and backscattering light collection restrict the majority of the backscattered photons to go through only one backscattering event. Therefore, higher order scattering events only broaden the probability depth distribution $p(z)$ compared to the purely double scattering events whereas the value of $z_{mp}$ remains approximately unchanged. However, for larger coherence lengths ($L_{sc} \sim l_s$; e.g. $L_{sc}$ = 90 μm), $z_{mp}$ obtained by the analytical model deviates from the one obtained by the numerical simulations due to emergent effect of higher order scattering events ($n_s$ > 5) and fails for $L_{sc} \gg l_s > l_s^*$. We conclude that in the low-coherence regime, which is the subject of our investigation, the analytical model enables accurate prediction of both *p(z)* and $z_{mp}$, and, thus, can be used to model *p(z)* and $z_{mp}$ of LEBS photons.

Figure 7 shows the dependence of $z_{mp}$ on the optical properties of a medium ($l_s$ and *g*) and the spatial coherence length of illumination $L_{sc}$ using the analytical model (Eq. (15)). The figures are plotted for different values of *g* (0.7 – 0.9) and $l_s$ (80 µm – 500 µm) for a constant $L_{sc}$ ($L_{sc}$ = 5 µm). As seen, $z_{mp}$ shows a relatively weak dependence on the optical properties of the medium when $L_{sc} < l_s << l_s^*$ (Figs. 7(a) and 7(b)). However, as shown in Fig. 6, $z_{mp}$ depends primarily on $L_{sc}$. This property of $z_{mp}$ is critical for LEBS measurements in biomedical applications as it enables probing a given physical depth of a biological tissue. That is, by adjusting the $L_{sc}$ of a light source, it should be possible to collect photons propagating into a tissue up to the depth of interest regardless of specific optical properties of a given tissue sample. It is also noticed that for a given $L_{sc}$, $z_{mp}(g,l_s) \propto \frac{g^{1/3}}{(1-g)^{2/3}}[l_s]^{1/3}$ which is a much slower varying function of *g* and $l_s$ than another length scale frequently used to describe light transport in tissue, $l_s^* = \frac{l_s}{(1-g)}$.

Experimental realizations to obtain the information about the depth of penetration of LEBS photons can be implemented in different ways as follows. a) The depth to which a photon penetrates can be experimentally estimated by varying the thickness of the sample. Varying the thickness provides a simple method for quantifying the contribution of different depths to the LEBS signal [34]. b) Time resolved measurements can also be used to assess the penetration depth by measuring short light pulses backscattered from the sample without any sample preparation. The depth of a photon inside the sample can then be experimentally gated based on the time of flight of such short light pulses [35]. We are currently considering a number of such experimental methodologies to implement our analytical derivations in potential experiments.

## IV. CONCLUSIONS

We have derived an analytical model of the penetration depth distribution *p(z)* and effective penetration depth $z_{mp}$ of photons that generate an LEBS peak (spatial coherence length $L_{sc} < l_s << l_s^*$), in the sub-diffusive scattering regime. We have performed numerical Monte Carlo simulations to support our analytical results. The results from the analytical model are in good agreement with those obtained from the Monte Carlo simulations. Our results demonstrate that $z_{mp} \propto (g^{1/3}/(1-g)^{2/3})[l_s \pi L_{sc}^2]^{1/3}$, *i.e.* $z_{mp}$ of the LEBS photon is approximately proportional to the 1/3 power of an effective coherence volume, $[l_s \pi L_{sc}^2]^{1/3}$ in an experimentally relevant parameter regime. More importantly, LEBS photons typically penetrate less than the scattering mean free path of the medium $l_s$ (that is, $z_{mp} < l_s$) when $L_{sc} < l_s << l_s^*$. Furthermore, the analytical calculation and numerical simulation show strong dependence of $z_{mp}$ on $L_{sc}$ (that can be controlled externally) and relatively weak dependence to tissue optical properties $(l_s, g)$, which suggests the possibility of using LEBS for depth-selective analysis of weakly scattering media such as biological tissue.

## Acknowledgements

This study was supported by National Health Institute grants R01CA112315, R01EB003682, and Coulter Foundation Award. Y.L. Kim was supported by a National Cancer Institute training grant R25 CA100600-01A1. P. Pradhan was supported by the Cancer Research Prevention Foundation (CRPF) and the Center for Cancer Nanotechnology Excellence (CCNE), Northwestern University, Evanston, IL.

**Figure Captions:**

**FIG 1:** "(Color online)" A schematic picture of a photon that undergo double scattering and exit within a very small radial distance $r$. $z'$ is the vertical distance of the first scatterer from the surface of the medium, $z''$ is the vertical distance of the second scatter and $\theta$ is the scattering angle. In order for the photon to undergo double scattering and exit within the narrow radial distance ($r < l_s << l_s^*$, $l_s$ - scattering mean free path of the medium, $l_s^* = \frac{l_s}{1-g}$, g – anisotropy factor), one of the scattering event occurs closer to the surface of the medium. That is $z' << z''$ or $z'' << z'$.

**FIG 2:** "(Color)" The scattering distribution of the photons $p(n_s)$ vs number of scattering $n_s$ from the numerical simulation for $r < l_s << l_s^*$ and $L_{sc} < l_s << l_s^*$ ($L_{sc}$ - spatial coherence length of illumination) for two different media with anisotropy factors g = 0.9 and 0.7 at constant $l_s$ = 100 μm. The photons predominantly undergo double scattering at small '$r$' and '$L_{sc}$'. However, the contribution from the double scattering photons decreases with increase in $r$ and $L_{sc}$.

**FIG 3:** "(Color)" The penetration depth distribution $p(z)$ of the photons vs depth $z$ for $r < l_s << l_s^*$ and $L_{sc} < l_s << l_s^*$ for two different media with anisotropy factors g = 0.9 and 0.7 at constant $l_s$ = 100 μm. The $p(z)$ of the photons predicted by the numerical simulations suggest a strong dependence on $r$ and $L_{sc}$ and relatively weak dependence on the optical properties $l_s$ and $g$.

**FIG 4:** "(Color online)" The penetration depth distribution $p(z)$ of single scattering photons plotted against those from double scattering photons at different depth $z$. At $r < l_s << l_s^*$, the first scatterer is located closer to the surface of the medium while the second scatterer is located several orders deeper than the first scatterer. Hence, the difference in vertical distances of the two scatters can be taken as the penetration depth of the photon.

**FIG 5:** "(Color)" Comparison of penetration depth distribution $p(z)$ of photons obtained from numerical simulation and analytical expression. (a) The distributions are obtained for a medium with g = 0.9 and $l_s$ = 100 μm for a radial distance $r$ = 5 μm and 25 μm. (b) The distributions are obtained for a medium with g = 0.9 and $l_s$ = 100 μm for a spatial coherence length, $L_{sc}$ = 5 μm and 25 μm.

**FIG 6:** "(Color)" Comparison between the effective penetration depth $z_{mp}$ of the photons that form a LEBS peak predicted by the numerical model and analytical expression (Eq. (15)). $z_{mp}$ is obtained as a function of $L_{sc}$ for two different media with anisotropy factors, g = 0.7 and g = 0.9 and a constant $l_s$ ($l_s$ = 100 μm). The agreement between the numerical model and analytical expression decreases with the increase in $L_{sc}$ as the photons undergo higher order scatterings.

**FIG 7:** "(Color online)" The dependence of $z_{mp}$ on individual optical properties plotted against $l_s$, and $g$. (a) $z_{mp}$ increases slowly between the anisotropy values of 0.7 and 0.9 after which $z_{mp}$ increases sharply. (b) $z_{mp}$ depends on $l_s$ only slightly over the range 100 μm and 500 μm that is relevant to the biological systems.

**Figures:**

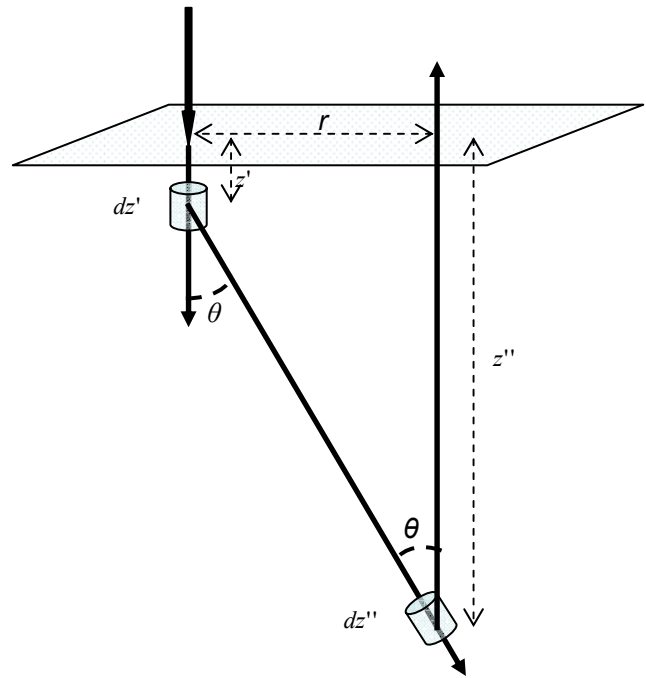

**FIG 1**

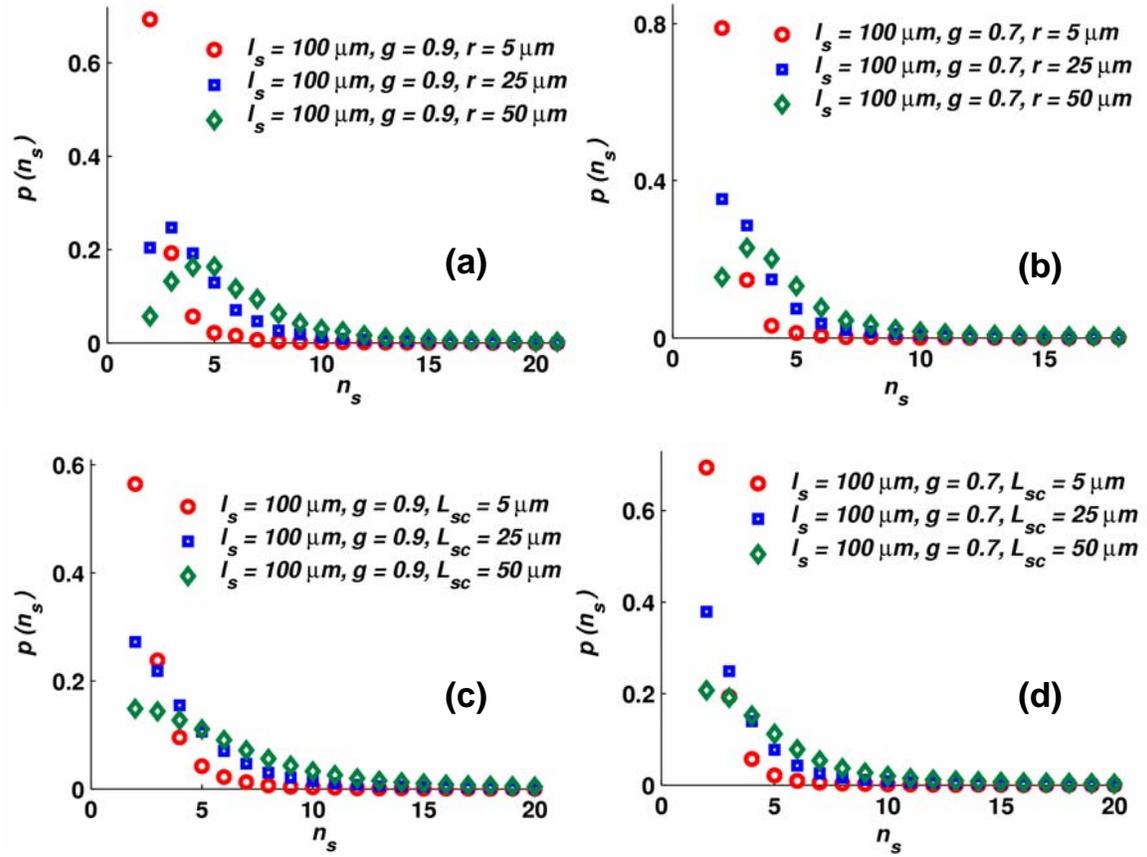

**FIG 2**

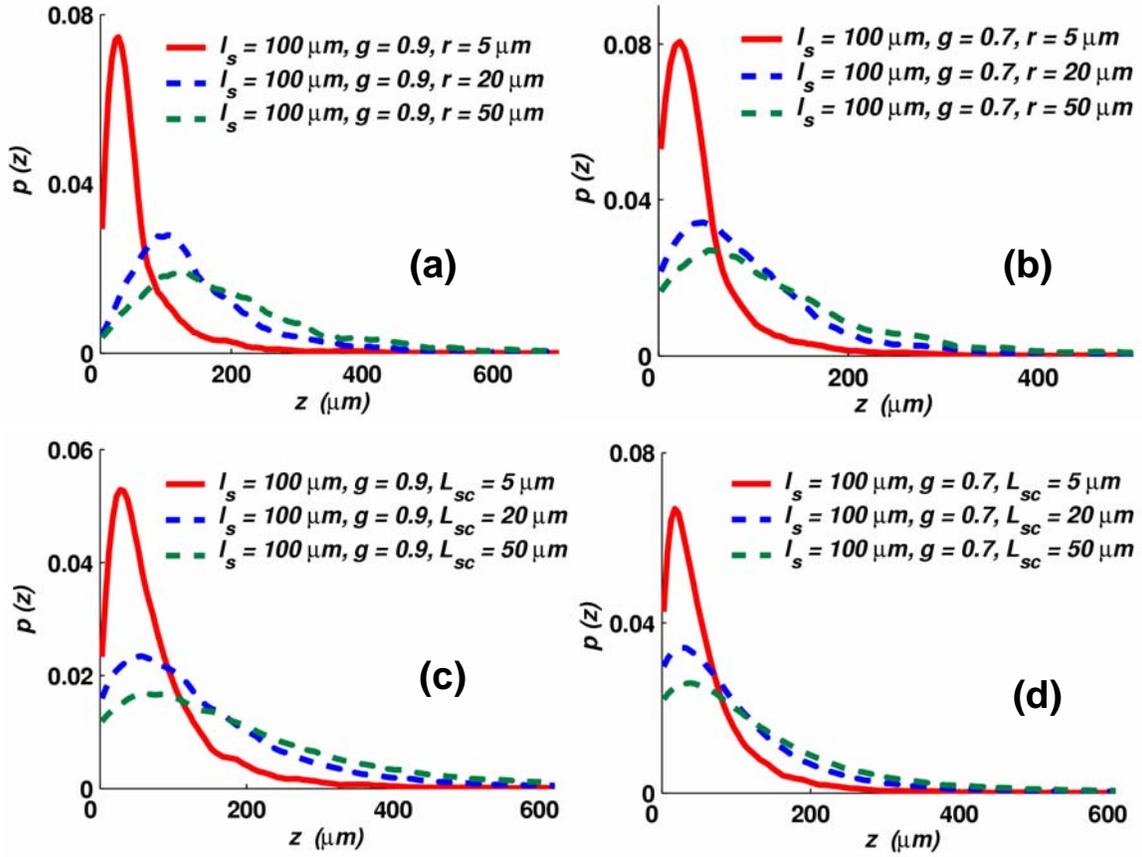

**FIG 3**

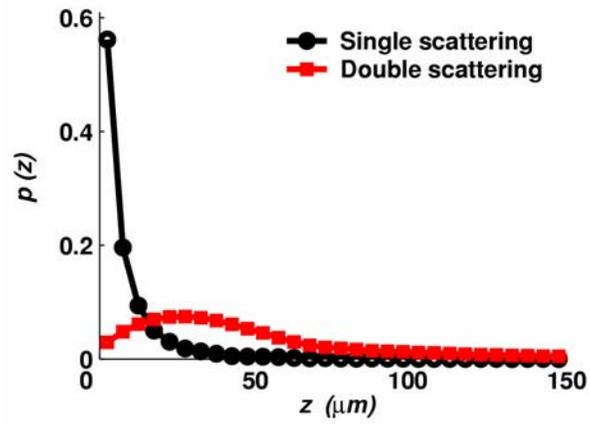

**FIG 4**

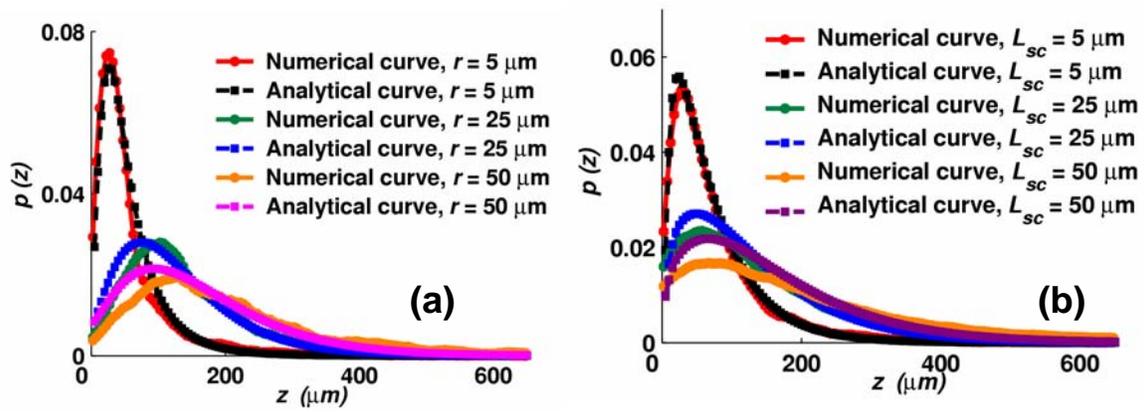

**FIG 5**

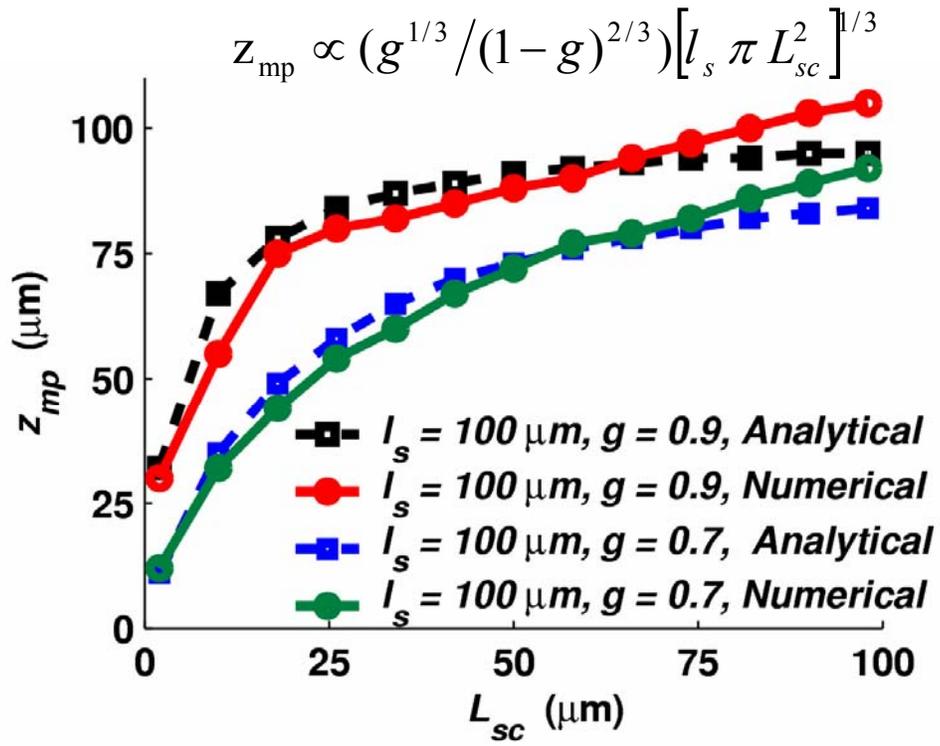

**FIG 6**

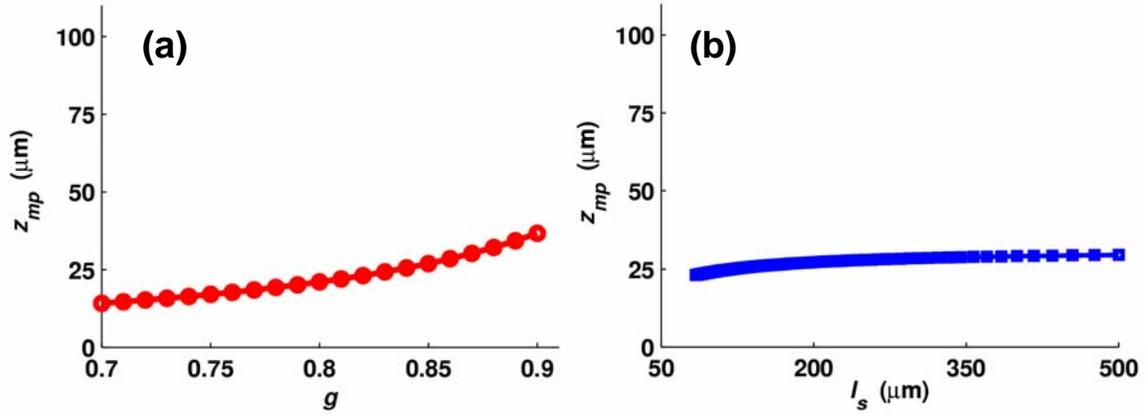

**FIG 7**